\documentclass[aps]{revtex4}
\usepackage{graphicx}

\begin{document}

\newcommand{\eq}{eqnarray}
\newcommand{\La}{{\Lambda}}
\newcommand{\om}{{\omega}}
\newcommand{\ra}{\rightarrow}

\title{Solar System tests of Ho\v{r}ava-Lifshitz black holes}

\author{Francisco S. N. Lobo}
\email{flobo@cii.fc.ul.pt} \affiliation{Centro de F\'{i}sica
Te\'{o}rica e Computacional, Faculdade de Ci\^{e}ncias da
Universidade de Lisboa, Campo Grande, Ed. C8 1749-016 Lisboa,
Portugal}

\author{Tiberiu Harko}
\email{harko@hkucc.hku.hk} \affiliation{Department of Physics and
Center for Theoretical and Computational Physics, The University
of Hong Kong, Pok Fu Lam Road, Hong Kong}

\author{Zolt\'{a}n Kov\'{a}cs}
\email{zkovacs@mpifr-bonn.mpg.de} \affiliation{Department of
Physics and Center for Theoretical and Computational Physics, The
University of Hong Kong, Pok Fu Lam Road, Hong Kong}

\date{\today}

\begin{abstract}

In the present paper
we consider the possibility of observationally testing Ho\v{r}ava
gravity at the scale of the Solar System, by considering the
classical tests of general relativity (perihelion precession of
the planet Mercury, deflection of light by the Sun and the radar
echo delay) for the Kehagias-Sfetsos asymptotically flat black hole solution. All these gravitational effects can
be fully explained in the framework of the vacuum solution of
Ho\v{r}ava gravity, and it is shown that the analysis of the classical general
relativistic tests severely constrain the free parameter of the
solution.

\end{abstract}

\maketitle


\section{Introduction}

Recently, Ho\v{r}ava proposed a renormalizable gravity theory in four dimensions
which reduces to Einstein gravity with a non-vanishing
cosmological constant in IR but with improved UV behaviors \cite {Horava:2008ih,Horava:2009uw}. The latter theory admits a Lifshitz scale-invariance in time and space, $t \rightarrow l^z\,t$ and $x^i \rightarrow l\,x^i$ (in particular, $z=3$ for the present case), exhibiting a broken Lorentz symmetry
at short scales, while at large distances higher derivative terms
do not contribute, and the theory reduces to standard general
relativity (GR).
Since one of the most important aspects of the theory is a Lifshitz-type anisotropic scaling, it is often called Ho\v{r}ava-Lifshitz gravity.
The Ho\v{r}ava theory has
received a great deal of attention and since its formulation
various properties and characteristics have been extensively
analyzed, ranging from formal developments \cite{formal},
cosmology \cite{cosmology}, dark energy \cite{darkenergy} and dark
matter \cite{darkmatter}, and spherically symmetric or rotating
solutions
\cite{BHsolutions,Park:2009zra,Lu:2009em,Kehagias:2009is, ghodsi}.

The natural setting of Ho\v{r}ava-Lifshitz gravity is the ADM formalism, where the 4-dim metric is parameterized by the
following
\begin{equation}
ds^{2}=-N^{2}c^{2}dt^{2}+g_{ij}\left( dx^{i}+N^{i}\,dt\right)
\left( dx^{j}+N^{j}\,dt\right) \,.
\end{equation}
$N$ is the lapse function and $N_i$ the shift function, respectively.

The IR-modified Ho\v{r}ava action is given by
\begin{eqnarray}
S &=&\int dt\,d^{3}x\;\sqrt{g}\,N\Bigg[\frac{2}{\kappa ^{2}}\left(
K_{ij}K^{ij}-\lambda _{g}K^{2}\right) -\frac{\kappa ^{2}}{2\nu
_{g}^{4}} C_{ij}C^{ij}+\frac{\kappa ^{2}\mu }{2\nu
_{g}^{2}}\epsilon ^{ijk}R_{il}^{(3)}\nabla _{j}R^{(3)l}{}_{k}
   \nonumber \\
&&-\frac{\kappa ^{2}\mu ^{2}}{8}R_{ij}^{(3)}R^{(3)ij}
+\frac{\kappa ^{2}\mu ^{2}}{8(3\lambda _{g}-1)}\left(
\frac{4\lambda _{g}-1 }{4}(R^{(3)})^{2}-\Lambda
_{W}R^{(3)}+3\Lambda _{W}^{2}\right)
   +\frac{\kappa ^{2}\mu ^{2}\omega }{8(3\lambda _{g}-1)}R^{(3)}\Bigg],
\label{Haction}
\end{eqnarray}
where $\kappa $, $\lambda _{g}$, $\nu _{g}$, $\mu $, $\omega $ and
$\Lambda _{W}$ are constant parameters.
$R^{(3)}$ is the three-dimensional curvature scalar for $g_{ij}$;
$K_{ij}$ is the extrinsic curvature given by
\begin{equation}
K_{ij}=\frac{1}{2N}\left( \dot{g}_{ij}-\nabla _{i}N_{j}-\nabla
_{j}N_{i}\right) \,,
\end{equation}
and $C^{ij}$ is the Cotton-York tensor, defined as
\begin{equation}
C^{ij}=\epsilon ^{ikl}\nabla _{k}\left( R^{(3)j}{}_{l}-\frac{1}{4}
R^{(3)}\delta _{l}^{j}\right) .
\end{equation}

The fundamental constants of the speed of light $c$, Newton's
constant $G$, and the cosmological constant $\Lambda$ are provided by the parameters of the theory and are defined as
\begin{equation}
c^2=\frac{\kappa^2\mu^2|\Lambda_W|}{8(3\lambda_g-1)^2},\quad
G=\frac{ \kappa^2c^2}{16\pi(3\lambda_g-1)},\quad
\Lambda=\frac{3}{2}\Lambda_W c^2,
\end{equation}
respectively.

There are basically four versions of Ho\v{r}ava gravity, namely, those
with or without the ``detailed balance condition'', and with or without the ``projectability condition''.
The ``detailed balance condition'' restricts the form of the potential in the 4--dim Lorentzian action to a specific form in terms of a 3--dim Euclidean theory. In a cosmological context, this condition leads to obstacles, and
thus must be abandoned.
In this context, the last term in Eq.~(\ref{Haction}) represents a `soft'
violation of the `detailed balance' condition, which modifies the
IR behavior. This IR modification term, $\mu ^{4}R^{(3)}$, with an
arbitrary cosmological constant, represent the analogs of the
standard Schwarzschild-(A)dS solutions, which were absent in the
original Ho\v{r}ava model.
The ``projectability condition'' essentially stems from the fundamental symmetry of the theory, i.e., the foliation-preserving diffeomorphism invariance, and must be
respected. Foliation-preserving diffeomorphism consists of a 3--dim spatial
and the space-independent time reparametrization. Note that the lapse function is essentially the gauge degree of freedom associated with the time reparametrization, so that it is natural to restrict it to be space-independent, i.e., $N=N(t)$.

IR-modified Ho\v{r}ava gravity seems to be
consistent with the current observational data, but in order to
test its viability more observational constraints are necessary \cite{Harko:2009qr,solarsystem}.
Thus, it is the purpose of the present paper to consider the classical
tests (perihelion precession, light bending, and the radar echo
delay, respectively) of general relativity for static
gravitational fields in the framework of Ho\v{r}ava-Lifshitz
gravity. To do this we shall adopt for the geometry outside a
compact, stellar type object (the Sun), the static and spherically
symmetric metric obtained by Kehagias and Sfetsos \cite{Kehagias:2009is}. This work is essentially based on the paper \cite{Harko:2009qr}.

\section{Static and spherically symmetric black hole solutions}

Consider a static and spherically symmetric solution, with the metric
ansatz:
\begin{equation}
ds^{2}=-e^{\nu (r)}dt^{2}+e^{\lambda (r)}dr^{2}+r^{2}\,(d\theta
^{2} +\sin^{2}{\theta }\,d\phi ^{2})\,.  \label{SSSmetric}
\end{equation}
By substituting the metric into the action, the resulting reduced
Lagrangian, after angular integration, is given by
\begin{\eq}
{\cal{L}}&=&\frac{\kappa^2\mu^2}{8(1-3\lambda_g)}e^{(\nu+\lambda)/2}
\left[(2\lambda_g-1)\frac{(e^{-\lambda}-1)^2}{r^2}
+2\lambda_g\frac{e^{-\lambda}-1}{r}(\lambda'e^{-\lambda})
+\frac{\lambda_g-1}{2}(\lambda'e^{-\lambda})^2\right.\nonumber
\\ &&\left. ~~-2 (\om-\La_W) (1-e^{-\lambda}(1-r\lambda'))
- 3 \La_W^2 r^2 \right]\ .
\end{\eq}

The above reduced Lagrangian yields the following equations of motions
\begin{eqnarray}
(2\lambda_g-1)\frac{(e^{-\lambda}-1)^2}{r^2}-
2\lambda_g\frac{e^{-\lambda}-1}{r}(-\lambda'
e^{-\lambda})+\frac{\lambda_g-1}{2}(\lambda' e^{-\lambda})^2
-2 (\om-\La_W) [1-e^{-\lambda}(1-r\lambda')]- 3 \La_W^2 r^2 &=&0\,,
\\
\frac{(\nu'+\lambda')}{2} \left[(\lambda_g-1)(\lambda'
e^{-\lambda})-2\lambda_g \frac{e^{-\lambda}-1}{r}+2(\om-\La_W)
r\right]
+(\lambda_g-1) \left[(-\lambda''+(\lambda)')e^{-\lambda}
-\frac{2(e^{-\lambda}-1)}{r^2}\right]&=&0\,,
\end{eqnarray}
by varying the functions $\nu$ and $\lambda$, respectively.

Now, imposing $\lambda _g=1$, which reduces to the
Einstein-Hilbert action in the IR limit, one obtains the following
solution of the vacuum field equations in Ho\v{r}ava gravity,
\begin{equation}
e^{\nu(r)}=e^{-\lambda(r)}=1+(\omega-\Lambda_W)r^2-\sqrt{
r[\omega(\omega-2\Lambda_W)r^3 +\beta]},  \label{gensolution}
\end{equation}
where $\beta$ is an integration constant. Throughout this work, we consider the Kehagias-Sfetsos (KS)
asymptotically flat solution \cite{Kehagias:2009is}, i.e., $\beta=4\omega M$ and
$\Lambda_W=0$:
\begin{equation}
e^{\nu(r)}=1+\omega r^2-\omega r^2\sqrt{1+\frac{4M}{\omega r^3}}.
\label{KSsolution}
\end{equation}

If the limit $4M/\omega r^3\ll 1$, from Eq.~(\ref{KSsolution}) we
obtain the standard Schwarzschild metric of general relativity,
$e^{\nu (r)}=1-2M/r$.
There are two event horizons at
$r_{\pm}=M\left[1\pm\sqrt{1-1/(2\omega M^2)}\right]$. To avoid a naked singularity at the origin, impose the condition
$\omega M^2\geq \frac{1}{2}$.
Note that in the GR regime, i.e., $\omega M^2 \gg 1$, the outer
horizon approaches the Schwarzschild horizon, $r_+\simeq 2M$, and
the inner horizon approaches the central singularity, $r_- \simeq
0$.

\section{Solar system tests for Ho\v{r}ava-Lifshitz black holes} \label{sect3}

There are three fundamental tests which provide observational evidence for
GR and its generalizations, namely, the precession of the perihelion of Mercury, the deflection of light by the Sun, and the radar echo delay observations. The latter have been used to
successfully test the Schwarzschild solution of general relativity
and some of its generalizations. In this Section, we consider these
standard Solar System tests of general relativity in the case of
the KS asymptotically flat solution \cite{Kehagias:2009is} of
Ho\v{r}ava-Lifshitz gravity. Throughout this Section we use
the natural system of units with $G=c=1$.
The analysis outlined below is essentially based on the paper \cite{Harko:2009qr}, and a similar analysis has been carried out in the context of branewold models \cite{Boehmer:2008zh}.

\subsection{The perihelion precession of Mercury}

The line element, given by Eq.~(\ref{SSSmetric}), provides the
following equation of motion for $r$
\begin{equation}
\dot{r}^{2}+e^{-\lambda }\frac{L^{2}}{r^{2}}=e^{-\lambda }\left(
E^{2}e^{-\nu }-1\right) \,,  \label{inter1}
\end{equation}
where
$e^{\nu }\dot{t}=E=\mathrm{constant}$ and $r^{2}\dot{\phi}=L=\mathrm{constant}$.

The analysis below is simplified by applying a change of variable $r=1/u$, and using $d/ds=Lu^{2}d/d\phi $. We further formally represent $e^{-\lambda }=1-f(u)$, so that the equation of motion (\ref{inter1}) takes the following form
\begin{equation}
\frac{d^{2}u}{d\phi ^{2}}+u=F(u),
\label{inter2}
\end{equation}
where $F(u)=\frac{1}{2}\frac{dG(u)}{du}$, and $G(u)$ is defined as
\begin{equation}
G(u)\equiv f(u)u^{2}+\frac{E^{2}}{L^{2}} e^{-\nu -\lambda
}-\frac{1}{L^{2}}e^{-\lambda }. \label{ueq_basic}
\end{equation}

A circular orbit $u=u_{0}$ is given by the root of the equation $
u_{0}=F\left( u_{0}\right) $. Any deviation $\delta =u-u_{0}$ from
a circular orbit must satisfy the equation
\begin{equation}
\frac{d^{2}\delta }{d\phi ^{2}}+\left[ 1-\left( \frac{dF}{du}\right)
_{u=u_{0}}\right] \delta =O\left( \delta ^{2}\right) ,
\end{equation}
which is obtained by substituting $u=u_{0}+\delta $ into Eq.
(\ref{inter2}). Therefore, in the first order in $\delta $, the
trajectory is given by
\begin{equation}
\delta =\delta _{0}\cos \left( \sqrt{1-\left( \frac{dF}{du}\right)
_{u=u_{0}} }\phi +\beta \right) ,
\end{equation}
where $\delta _{0}$ and $\beta $ are constants of integration.

The variation of the orbital angle from one perihelion to the next
is $\phi =2\pi/(1-\sigma)$, where $\sigma$ is the perihelion advance, which represents the rate of
advance of the perihelion. $\sigma$ is given by $\sigma
=1-\sqrt{1-\left(dF/du\right) _{u=u_{0}}}$, or for small $\left(
dF/du\right) _{u=u_{0}}$, by
$\sigma =\frac{1}{2}\left( \frac{dF}{du}\right) _{u=u_{0}}$. As the planet advances $\phi $ radians in its orbit, its perihelion advances $\sigma \phi $ radians.
For a complete rotation we have $\phi \approx 2\pi (1+\sigma )$,
and the advance of the perihelion is
$\delta \phi =\phi -2\pi\approx 2\pi \sigma$.

\begin{figure}[h]
\includegraphics[width=3.05in]{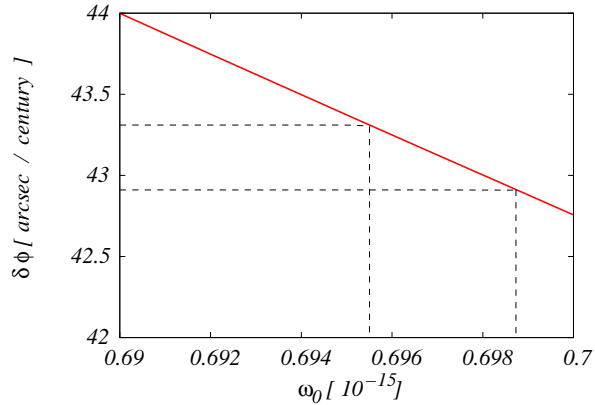}
\caption{Variation of the precession angle $\delta \phi$ as a
function of $\omega _0$.} \label{fig2}
\end{figure}

For the KS solution, the perihelion precession is given by:
\begin{equation}
\delta \phi =\pi\frac{3 \sqrt{\omega _0 } \left\{2
\left(x^3_0+\omega _0 \right) x^5_0+b^2 \left[2 x^6_0 +\left(6
\omega _0 -4 \sqrt{\omega _0  \left(4
   x^3_0+\omega _0 \right)}\right) x^3_0+\omega _0 ^2
   -\sqrt{\omega _0 ^3 \left(4 x^3_0+\omega _0
   \right)}\right]\right\}}{x^4_0 \left(4
   x^3_0+\omega _0 \right)^{3/2}}.
\end{equation}
with the dimensionless parameters defined as $\omega_0=M^2\omega$, and $x_{0}=u_{0}M$.
In the ``Post-Newtonian'' limit $4x_0^3/\omega _0\ll 1$, we obtain
the GR result $\delta \phi _{GR}=6\pi b^2$, where $b^{2}=M/a\left(
1-e^{2}\right)$. The variation of the perihelion precession angle as a
function of $\omega _0$ is represented in Fig.~\ref{fig2}.

The observed value of the perihelion precession of the planet
Mercury is $\delta \phi _{Obs}=43.11\pm 0.21$ arcsec per century \cite{Sh}.
Therefore the range of variation of the perihelion precession is
$\delta \phi _{Obs}\in \left(42.90,43.32\right)$.
This range of observational values fixes the range of $\omega _0$
as
\begin{equation}
\omega _0\in\left(6.95431\times 10^{-16},6.98821\times
10^{-16}\right).
\end{equation}

\subsection{The deflection of light}

The deflection angle of a light ray by the gravitational field of
a massive body in a spherically symmetric geometry is derived in \cite{Wein}, and is given by
$\Delta \phi =2\left| \phi \left( r_{0}\right) -\phi \left( \infty \right)
\right| -\pi $,
with
\begin{equation}
\phi \left( r\right) =\phi \left( \infty \right) +\int_{r}^{\infty
}\frac{ e^{\lambda /2}}{\sqrt{e^{\nu \left( r_{0}\right) -\nu
\left( r\right) }\left( r/r_{0}\right) ^{2}-1}}\frac{dr}{r},
\end{equation}
where $r_{0}$ is the distance of the closest approach. Here we
have taken into account that in the absence of a gravitational
field a light ray propagates along a straight line. In the
case of the Sun, the deflection angle of alight ray is $\Delta
\phi =1.72752^{\prime \prime }$.

The deflection angle of light rays passing nearby the Sun in the
KS geometry is given by
\begin{equation}
\phi \left( x_{0}\right) =\phi \left( \infty \right)
+\int_{1}^{\infty } \frac{\left[ 1+\omega
_{0}x_{0}^{2}x^{2}-\sqrt{x_{0}x(\omega
_{0}^{2}x_{0}^{3}x^{3}+4\omega _{0})}\right] ^{-1/2}}{\sqrt{\left[
1+\omega _{0}x_{0}^{2}-\sqrt{x_{0}(\omega
_{0}^{2}x_{0}^{3}+4\omega _{0})}\right] x^{2}/\left[ 1+\omega
_{0}x_0^2x^{2}-\sqrt{x_{0}x(\omega _{0}^{2}x_{0}^{3}x^{3}+4\omega
_{0})}\right] -1}}\frac{dx}{x},
\end{equation}
where $\omega =\omega _{0}/M^{2}$, $ r_{0}=x_{0}M$, and
considering $r=r_{0}x$.

\begin{figure}[h]
\includegraphics[width=3.05in]{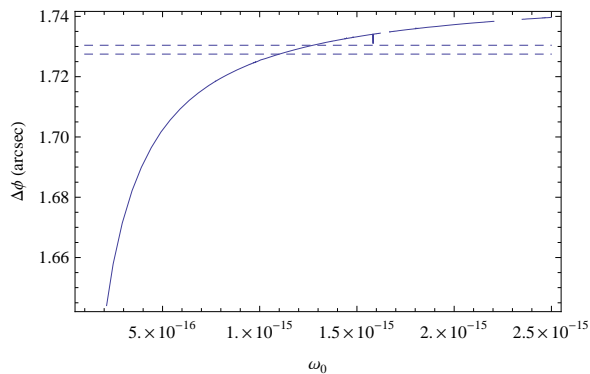}
\caption{The light deflection angle $\Delta \phi $ (in arcseconds)
as a function of the parameter $\omega _0$.} \label{fig3}
\end{figure}

For the Sun, by taking $r_{0}=R_{\odot }=6.955\times 10^{10}$ cm,
we find for $x_{0}$ the value $x_{0}=4.71194\times 10^{5}$. The
variation of the deflection angle $\Delta \phi =2\left| \phi
\left( x_{0}\right) -\phi \left( \infty \right) \right| -\pi $ is
represented, as a function of $\omega _{0}$, in Fig.~\ref{fig3}.

The best available data come from long baseline radio
interferometry \cite{all2}, which gives $\delta \phi _{LD}=\delta \phi
_{LD}^{(GR)}\left( 1+\Delta _{LD}\right) $, with $\Delta _{LD}\leq
0.0017$, where $\delta \phi _{LD}^{(GR)}=1.7275 $ arcsec. Thus, the observational constraints of light deflection restricts the
value of $\omega _0$ to
\begin{equation}
\omega _0\in \left(1.1\times10^{-15},1.3\times 10^{-15}\right)\,.
\end{equation}

\subsection{Radar echo delay}

The aim of this test is to measure the time required for radar signals to travel to an inner planet or satellite in two circumstances, namely,
(i) when the signal passes very near the Sun and, (ii) when the ray does not go near the Sun.
The time of travel of light $t_{0}$ between two planets, situated
far away from the Sun, is given by
$t_{0}=\int_{-l_{1}}^{l_{2}}dy$, where $l_{1}$ and $l_{2}$ are the
distances of the planets to the Sun, respectively.

If the light travels close to the Sun, the time travel is
\begin{equation}
t=\int_{-l_{1}}^{l_{2}}\frac{dy}{v}=\int_{-l_{1}}^{l_{2}}e^{\left[\lambda
(r)-\nu (r)\right] /2}dy,
\end{equation}
and the time difference is given by
\begin{equation}
\Delta t=t-t_{0}=\int_{-l_{1}}^{l_{2}}\left\{ e^{\left[ \lambda
(r) -\nu (r) \right] /2}-1\right\} dy=\int_{-l_{1}}^{l_{2}}\left\{
e^{\left[ \lambda \left( \sqrt{ y^{2}+R_{\odot }^{2}}\right) -\nu
\left( \sqrt{y^{2}+R_{\odot }^{2}}\right) \right] /2}-1\right\}
dy,
\end{equation}
with $r=\sqrt{y^{2}+R_{\odot }^{2}}$, and $R_{\odot }$ is the
radius of the Sun.

Recent measurements of the frequency
shift of radio photons to and from the Cassini spacecraft as they
passed near the Sun have greatly improved the observational
constraints on the radio echo delay. For the time delay of the signals emitted on Earth, and which graze the Sun, one obtains $\Delta t_{RD}=\Delta
t_{RD}^{(GR)}\left( 1+\Delta _{RD}\right) $, with $\Delta
_{RD}\leq \left(2.1\pm2.3\right)\times 10^{-5}$ \cite{Re79}.
The standard GR radar echo delay value is $\Delta
t_{RD}^{(GR)}\approx 4M_{\odot}\ln \left( 4l_{1}l_{2}/R_{\odot
}^{2}\right) \approx 2.4927\times 10^{-4}\, \mathrm{s}$.

\begin{figure}[h]
\includegraphics[width=3.05in]{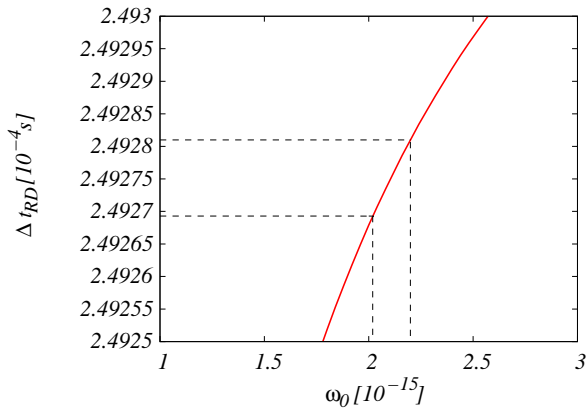}
\caption{Variation of the time delay $\Delta t_{RD}$ as a function
of $\omega _0$.} \label{fig4}
\end{figure}

For the KS solution, by introducing a new variable $\xi $ defined as $y=2\xi
M_{\odot}$, and by representing $\omega $ as $\omega =\omega
_{0}/M_{\odot}^{2}$, we obtain for the time delay the following expression
\begin{equation}
\Delta t_{RD}=16M_{\odot}\omega _{0}\int_{-\xi _{1}}^{\xi
_{2}}\frac{ \left( \xi ^{2}+a^{2}\right) \left[ \sqrt{1+\left(
1/2\omega _{0}\right) \left( \xi ^{2}+a^{2}\right)
^{-3/2}}-1\right] }{1-4\omega _{0}\left( \xi ^{2}+a^{2}\right)
\left[ \sqrt{1+\left( 1/2\omega _{0}\right) \left( \xi
^{2}+a^{2}\right) ^{-3/2}}-1\right] }d\xi ,
\end{equation}
where $a^{2}=R_{\odot }^{2}/4M_{\odot }^{2}$, $\xi
_{1}=l_{1}/2M_{\odot }$, and $\xi _{2}=l_{2}/2M_{\odot }$,
respectively. The variation of the time delay as a function of
$\omega _0$ is represented in Fig.~\ref{fig4}.

The observational values of the radar echo delay are consistent
with the KS black hole solution in Ho\v{r}ava-Lifshitz gravity if
\begin{equation}
\omega _0\in\left(2\times 10^{-15},3\times
10^{-15}\right) .
\end{equation}

\section{Conclusion}\label{sect5}

In the present work, we have considered the observational and
experimental possibilities for testing, at the level of the Solar
System, the Kehagias and Sfetsos solution of the vacuum field
equations in Ho\v{r}ava-Lifshitz gravity. In this context,
the parameter $\omega $, having the physical dimensions of ${\rm
length}^{-2}$, is constrained by the perihelion precession of
Mercury to a value of
$\omega =7\times 10^{-16}/M_{\odot }^2=3.212\times 10^{-26}$
cm$^{-2}$.
The deflection angle of the light rays by the Sun can be fully
explained in Ho\v{r}ava-Lifshitz gravity with the parameter
$\omega $ having the value
$\omega = 10^{-15}/M_{\odot }^2=4.5899\times 10^{-26}$,
while the radar echo delay experiment suggests a value of
$\omega =2\times 10^{-15}/M_{\odot }^2=9.1798\times 10^{-26}$
cm$^{-2}$.
From these values we can give an estimate of $\omega $ as
\begin{equation}
\omega =\left(5.660\pm3.1\right)\times 10^{-26}\;{\rm cm}^{-2}.
\end{equation}
This requires a very precise fine tuning of this constant at the level
of the Solar System.

Thus, the study of the classical tests of general relativity
provide a very powerful method for constraining the allowed
parameter space of Ho\v{r}ava-Lifshitz gravity solutions, and to
provide a deeper insight into the physical nature and properties
of the corresponding spacetime metrics.

\end{document}